\documentclass[11pt]{article}

\usepackage{amssymb}

\makeatletter
  \renewcommand{\theequation}{\thesection.\arabic{equation}}
  \@addtoreset{equation}{section}
\makeatother

\begin{document}
\setlength{\baselineskip}{6mm}

\begin{titlepage}

\begin{flushright}
KEK-TH-834 \\
ICRR-Report-491-2002-9\\
August 2002
\end{flushright}

%%% Title %%%

\vspace{10mm}
\begin{center}
{\Large \bf Comments on Quantum Aspects of Three-Dimensional} \\
{\Large \bf  de Sitter Gravity}

\vspace{10mm}

\renewcommand{\thefootnote}{\fnsymbol{footnote}}

{\large Hiroshi Umetsu}$^1$\footnote{
umetsu@post.kek.jp and umetsu@nbi.dk}  
{\large and Naoto Yokoi}$^2$\footnote{
yokoi@icrr.u-tokyo.ac.jp} 

\renewcommand{\thefootnote}{\arabic{footnote}}
\setcounter{footnote}{0}

\vspace{5mm}

{\it
$^1$ High Energy Accelerator Research Organization (KEK), \\
     Tsukuba, Ibaraki 305-0801, Japan \\
     and \\
     The Niels Bohr Institute, \\
     Blegdamsvej 17, DK-2100 Copenhagen \O, Denmark

\vspace{5mm}

$^2$ Institute for Cosmic Ray Research, University of Tokyo,\\
     Kashiwa, Chiba 277-8582, Japan
}
\end{center}

\vspace{5mm}

\begin{abstract}

We investigate the quantum aspects of three-dimensional gravity with a
positive cosmological constant. The reduced phase space of the 
three-dimensional de Sitter gravity is obtained as the space 
which consists of the Kerr-de Sitter space-times and 
their Virasoro deformations.
A quantization of the phase space
is carried out by the geometric quantization of 
the coadjoint orbits of the asymptotic Virasoro symmetries.
The Virasoro algebras with real central charges are obtained
as the quantum asymptotic symmetries. 
The states of globally de Sitter and point
particle solutions become the primary states of the unitary
irreducible representations of the Virasoro algebras. 
It is shown that those states are perturbatively stable 
at the quantum level.
The Virasoro deformations of these solutions correspond to
the excited states in the unitary irreducible representations.
In view of the dS/CFT correspondence, we also study the relationship
between the Liouville field theory obtained by a reduction of 
the SL(2;$\mathbb{C}$) Chern-Simons theory and 
the three-dimensional gravity 
both classically and quantum mechanically.  
In the analyses of the both theories, the Kerr-de Sitter geometries
with nonzero angular momenta do not give the unitary representations
of the Virasoro algebras. 

\end{abstract}

\end{titlepage}

\section{Introduction}
Quantum gravity on de Sitter space-time attracts much interest
owing to not only conceptual problems \cite{Witten1} (see also
\cite{Bousso} and references therein) 
but also recent astronomical observations \cite{Riess, Perlmutter}.  
Recently it was proposed that the gravity with a positive cosmological
constant has a dual description by the Euclidean conformal field theory 
on the infinitely past or future boundary \cite{Strominger} 
(see also \cite{SSV}).
Since de Sitter space-time has no everywhere time-like Killing
vector, constructions of quantum field theories on that space 
could be non-trivial.
And it is known that the quantum field theories on de Sitter space 
has some peculiar features, e.g., the existence of various possible
vacua \cite{SSV, BMS}. 
To deeply understand this correspondence, proper investigations of 
quantum theories of de Sitter gravity are needed because 
one has no well-defined stringy example of de Sitter space-time.
Since in three dimensions a gauge theoretical description of 
the gravity is known \cite{Townsend, Witten2}, 
it seems that we can analyze the gravity 
at the quantum level\footnote{From the viewpoint of the dS/CFT
correspondence, the Chern-Simons gravity on three-dimensional 
de Sitter space-time is also studied 
in \cite{Klemm1, BBM1, Cunha1, GKS, Klemm2, BBM2, Cunha2}.
In more general settings including the de Sitter gravity case, a
correspondence between the three-dimensional Chern-Simons theory with
boundary and a conformal field theory was earlier discussed in
\cite{Park2} as a generalization of the AdS/CFT correspondence by using
the Dirac's canonical method.}.  

Our aim in this paper is to construct concretely a quantum theory of 
the three-dimensional gravity with a positive cosmological constant.
For this end, 
we use the similar procedures to those used in Ref.\cite{NUY}
to investigate a quantum anti-de Sitter gravity.
We first construct the phase space of this system by using  
the SL(2;$\mathbb{C}$) Chern-Simons description of the three-dimensional
gravity \cite{Witten4}.
The phase space consists of solutions of the equations of motion
under the boundary conditions that the space-time should 
approach to de Sitter space in the asymptotically past \cite{Strominger}. 
It includes the Kerr-de Sitter geometries \cite{Deser-Jackiw, Park}
as particular solutions. 
The Virasoro algebras which are reduced symmetries of the gauge symmetry 
act on this space by their coadjoint forms 
and their central charges are given by a well-known value 
$c=\frac{3l}{2G}$.
Each geometry corresponding to an element on the phase space 
can be represented as a non-compact region in SL(2;$\mathbb{C}$)/SU(1,1) 
and its metric is induced from the Killing metric of 
SL(2;$\mathbb{C}$)/SU(1,1).
The actions of the Virasoro algebras on the phase space can be 
interpreted as deformations of that region.
The family which consists of geometries connected with each other 
by the Virasoro deformations is identified with a product of 
the coadjoint orbits of the conformal groups labeled by mass and 
angular momentum of the Kerr-de Sitter geometry.
Since the conformal group is not globally well-defined, 
our discussion of the coadjoint orbit would be formal.
Thus we mainly concentrate on the infinitesimal deformations of 
the Kerr-de Sitter geometries and neighborhood of the elements 
corresponding to those Kerr-de Sitter geometries on the orbits.

In Section 3 we consider a quantization of the Virasoro deformations 
which are degrees of freedom of the gravitational system 
analyzed here.
In Ref.\cite{Park} it was shown that the Virasoro algebras 
with imaginary central charges were acquired as algebras 
of the conserved charges by a quantization of the corresponding 
Chern-Simons theory.
In the appendix B, we provide some descriptions which state 
that the Virasoro algebra with an imaginary central charge 
are also obtained by a different manner of quantization.
We would like to show that the unitary irreducible representations of 
the Virasoro algebra with a {\it real} central charge can be obtained 
by geometric quantization of the coadjoint orbit prescribed by 
an appropriate symplectic form on the orbit.
Here strict analyses will be restricted within the perturbative level 
under the condition $c\gg 1$.
We consider the unitary representations of the Virasoro algebra 
adopting the unitarity conditions $L_m^\dagger =L_{-m}$ for the Virasoro
generators which are usually used in conformal field theories.
Then we discuss the perturbative stability of vacua 
by regarding $L_0$ as the energy function.
The unitarity and the stability of vacua require that 
the conformal weights of the primary states should be real and
positive. It means in the sense of the three-dimensional gravity that 
only the point particle geometries, i.e., without angular momenta, 
and their Virasoro deformations can provide the unitary and stable 
quantum theory. The geometries with excess angles are forbidden.

It is known that the Chern-Simons theory on a manifold with boundaries 
reduces to the chiral WZNW theories on the boundaries \cite{Elitzur}.
In the case of the Chern-Simons gravity with (negative) positive 
cosmological constant, it is shown that the chiral WZNW theories 
furthermore reduce to the Liouville field theory under the asymptotic 
(anti-)de Sitter boundary condition \cite{CHD-BO, Klemm1}.
In Section 4 we see the explicit correspondence between 
three-dimensional geometries and solutions of the Liouville theory. 
The relationship of infinitesimal Virasoro deformations of 
the Kerr-de Sitter geometries and those of the Liouville theory 
are provided.
At the quantum level, the direct product of the primary states 
with appropriate conformal weights in the Liouville theory 
is identified with the state of the Kerr-de Sitter space-time.
The states corresponding to the deformed geometries are identified 
with the excited states which are obtained by acting the Virasoro
generators on the primary state.

Finally we provide some discussions in Section 5.

\section{Three-Dimensional Geometry with Positive Constant Curvature}
Solutions of the Einstein equation with a positive cosmological constant 
$1/l^2$ are geometries with a positive constant scalar curvature $6/l^2$.
All solutions are locally equivalent to each other 
because the three-dimensional gravity has no local degrees of freedom.
In this section, we would like to clarify what characterizes 
each solution.
And we provide a phase space of this system under the boundary conditions 
being asymptotically de Sitter space-time from the perspective of 
Chern-Simons gravity.
\subsection{On the Kerr-de Sitter Solution}
The Kerr-de Sitter solution is characterized by its ADM-like mass 
$M=-\frac{\mu}{8G}$ and angular momentum $J$.
In the Schwarzschild coordinates $(t, \ r, \ \phi)$, 
its metric $ds^2_{(\mu, J)}$ is given by 
\begin{eqnarray}
\label{KdS-metric}
ds^2_{(\mu, J)}=-N^2 (dt)^2 +N^{-2} (dr)^2
    +r^2 \left(d\phi+N^{\phi}dt\right)^2,
\end{eqnarray}
where 
\begin{equation}
 N^2=\mu-\frac{r^2}{l^2}+\frac{16G^2J^2}{r^2}, \hspace{10mm}
 N^\phi=\frac{4GJ}{r^2}.
\end{equation}
The space-time has the cosmological horizon at $r=r_+$,
\begin{equation}
 r_+=\frac{l}{2}\left( \sqrt{\mu+i\frac{8GJ}{l}}
                +\sqrt{\mu-i\frac{8GJ}{l}} ~\right),
\end{equation}
if $\mu \in \mathbb{R}$ for $J\neq 0$ or $\mu \in \mathbb{R}_{\geq 0}$ for $J=0$. 
In the case $J=0$, this metric represents a point particle sitting at
$r=0$~\cite{Deser-Jackiw}, 
\begin{equation}
\label{pt-particle}
 ds^2_{(\mu, 0)}=
   -\left(1-\frac{\tilde{r}^2}{l^2}\right)\left(d\tilde{t}\right)^2
       +\frac{\left(d\tilde{r}\right)^2}{1-\frac{\tilde{r}^2}{l^2}}
        +\tilde{r}^2\left(d\tilde{\phi}\right)^2,
\end{equation}
where we transformed the coordinates as follows,
\begin{equation}
 \tilde{t}=\sqrt{\mu} ~t, \hspace{10mm}
 \tilde{r}=\frac{r}{\sqrt{\mu}}, \hspace{10mm}
 \tilde{\phi}=\sqrt{\mu} ~\phi.
\end{equation}
We have to make the identification 
$\tilde{\phi} \sim \tilde{\phi}+2\pi \sqrt{\mu}$ 
because $\phi \sim \phi +2\pi$. 
Thus this geometry has a conical singularity at $r=0$ with a deficit
angle $2\pi\left(1-\sqrt{\mu}\right)$ if $0<\mu < 1$.
On the other hand, we do not have any physical interpretation 
for space-time for $\mu >1$ which has an excess angle 
$2\pi\left(\sqrt{\mu}-1\right)$.
In the case of $J=0$ and $\mu<0$, $N^2$ becomes negative 
in the coordinate system used in the eq.(\ref{KdS-metric}).
Thus we will omit this case in the most part in this paper 
unless explicitly mentioned, 
although this metric can be rewritten in the form of de Sitter
metric locally by an appropriate coordinate transformation.
\footnote{
In fact, although one can transform it into de Sitter metric like 
(\ref{pt-particle}), time-like coordinate becomes periodic 
in the resulting coordinate system.}

We will mainly use another coordinate system $(\tau, \ z, \ \bar{z})$ 
which is defined below. 
By the coordinate transformations, 
\begin{eqnarray}
\label{coord-trans}
\tau&=&\frac{t}{l}+f(r)
    -\frac{1}{2}\ln\frac{\sqrt{b_0\bar{b}_0}}{c/6}, \nonumber \\ 
z&=&e^{\frac{t}{l}+g(r)+i\phi}, \\
\bar{z}&=&e^{\frac{t}{l}+\bar{g}(r)-i\phi}, \nonumber
\end{eqnarray}
the Kerr-de Sitter metric $ds^2_{(\mu, J)}$ is written 
in the following form,
\begin{equation}
\label{metric-b_0}
ds^2_{(b_0, \bar{b}_0)}=l^2\left[- d\tau^2 +\frac{b_0}{c/6}\frac{dz^2}{z^2}
                +\frac{\bar{b}_0}{c/6}\frac{d\bar{z}^2}{\bar{z}^2}
                +\left(e^{-2\tau}+\frac{b_0\bar{b}_0}{(c/6)^2 z^2 \bar{z}^2}
                  e^{2\tau}\right)dzd\bar{z} \right],
\end{equation}
where $c\equiv\frac{3l}{2G}$, and $b_0$, $\bar{b}_0$ are defined by 
\begin{equation}
\label{b_0}
b_0\equiv\frac{c}{24}(1-\mu)-\frac{i}{2}J, \hspace{10mm}
\bar{b}_0\equiv\frac{c}{24}(1-\mu)+\frac{i}{2}J.
\end{equation} 
And $f(r), g(r)$ and $\bar{g}(r)$ are obtained by solving the following 
first order differential equations, 
\begin{equation}
l\frac{d}{dr}f(r) = \frac{\sqrt{1-N^2}}{N^2}, \hspace{8mm}
l\frac{d}{dr}g(r) = \frac{\frac{r^2}{l^2}+\frac{b_0-\bar{b}_0}{c/6}}
                           {\frac{r^2}{l^2}N^2\sqrt{1-N^2}},
  \hspace{8mm}
l\frac{d}{dr}\bar{g}(r) = \frac{\frac{r^2}{l^2}-\frac{b_0-\bar{b}_0}{c/6}}
                           {\frac{r^2}{l^2}N^2\sqrt{1-N^2}}.
\end{equation}
For these coordinate transformations to be well-defined, 
the range of $r$ is restricted in
\begin{eqnarray}
0<r<l\sqrt{\mu}, && \hspace{10mm} 
    \mbox{for} \ \ 0\leq \mu \leq 1, \ J=0  \nonumber \\
l\sqrt{\mu-1}<r<l\sqrt{\mu}, && \hspace{10mm}
    \mbox{for} \ \ 1 < \mu, \ J=0 \\
l\left|\sqrt{\frac{b_0}{c/6}}-\sqrt{\frac{\bar{b}_0}{c/6}}\right|<r<r_+. 
    && \hspace{10mm} \mbox{for} \ \ \mu\in\mathbb{R}, \ J\neq 0 
     \nonumber
\end{eqnarray}
 Therefore we will consider the region of the Kerr-de Sitter 
space-time defined by $-\infty<t<\infty$, $0\leq\phi<2\pi$ 
and the above range of $r$. 

These geometries can be represented in the coset space 
SL(2;$\mathbb{C}$)/SU(1,1).\footnote{
SL(2;$\mathbb{C}$) generators we use in this paper are 
\begin{eqnarray*}
 J^0=\left(\begin{array}{cc}
      \frac{i}{2} & 0 \\
      0 & -\frac{i}{2}
	   \end{array}\right), \hspace{10mm}
 J^1=\left(\begin{array}{cc}
      0 & -\frac{1}{2} \\
      -\frac{1}{2} & 0
	   \end{array}\right), \hspace{10mm}
 J^2=\left(\begin{array}{cc}
      0 & -\frac{i}{2} \\
      \frac{i}{2} & 0
	   \end{array}\right).
\end{eqnarray*}
These satisfy $\left[J^a, \ J^b\right]=\epsilon^{abc} J_c$, 
where $\epsilon^{abc}$ is a totally anti-symmetric tensor 
$\epsilon^{012}=+1$ and {\it sl}(2;$\mathbb{C}$) indices are lowered 
(or raised) by $\eta_{ab}\equiv \textrm{diag.}(-1, \ 1, \ 1)$ 
(or $\eta^{ab}$, the inverse matrix of $\eta_{ab}$).
And $\left(J^a\right)^\dagger = J_a$.
}
Let us consider a non-compact region in SL(2;$\mathbb{C}$)/SU(1,1) 
consisting of the elements 
\begin{eqnarray}
\label{cal-G_0}
{\cal G}_0\equiv
 \left(\begin{array}{cc}
  \frac{a+1}{2\sqrt{a}}z^{\frac{a-1}{2}} & 
    -\frac{i}{\sqrt{a}}z^{\frac{a+1}{2}} \\
  -\frac{i(a-1)}{2\sqrt{a}}z^{-\frac{a+1}{2}} & 
    \frac{1}{\sqrt{a}}z^{-\frac{a-1}{2}}
       \end{array}\right)
 \left(\begin{array}{cc}
  e^{\tau} &  0 \\
  0 & e^{-\tau}
       \end{array}\right)
 \left(\begin{array}{cc}
  \frac{\bar{a}+1}{2\sqrt{\bar{a}}}\bar{z}^{\frac{\bar{a}-1}{2}} & 
    -\frac{i(\bar{a}-1)}{2\sqrt{\bar{a}}}\bar{z}^{-\frac{\bar{a}+1}{2}} \\
  -\frac{i}{\sqrt{\bar{a}}}\bar{z}^{\frac{\bar{a}+1}{2}} & 
    \frac{1}{\sqrt{\bar{a}}}\bar{z}^{-\frac{\bar{a}-1}{2}}
       \end{array}\right),
\end{eqnarray} 
where $a=\sqrt{1-\frac{b_0}{c/24}}$ and 
$\bar{a}=\sqrt{1-\frac{\bar{b}_0}{c/24}}$ which coincide with 
$\sqrt{\mu}$ for $J=0$.
${\cal G}_0$ is characterized by the relation 
${\cal G}_0=\epsilon {\cal G}_0^\dagger\epsilon^{-1}$ 
where $\epsilon\equiv2J^0$.
The metric on this region induced from the Killing metric of 
SL(2;$\mathbb{C}$)/SU(1,1) coincides with the Kerr-de Sitter metric 
(\ref{metric-b_0}),
\begin{equation}
ds^2_{(b_0, \bar{b}_0)}=
  -\frac{l^2}{2}\mbox{Tr}\left({\cal G}_0^{-1}d{\cal G}_0\right)^2.
\end{equation}
Therefore we can make the following interpretation: 
${\cal G}_0$ gives a mapping from the parameter space 
$(\tau, \ z, \ \bar{z})$ to SL(2;$\mathbb{C}$)/SU(1,1).
Then the Kerr-de Sitter space-time is realized as a region 
defined by ${\cal G}_0$ in SL(2;$\mathbb{C}$)/SU(1,1). 
Thus each solution is characterized by its region.
And its metric is induced from the Killing metric of 
SL(2;$\mathbb{C}$)/SU(1,1).

\subsection{SL(2;$\mathbb{C}$) Chern-Simons Gauge Theory}
It is well-known that the Chern-Simons gauge theory with an appropriate 
gauge group is on-shell equivalent to the three-dimensional
gravity\cite{Townsend, Witten2}.\footnote{
Of course, this equivalence holds only under some assumptions, 
e.g. the reversibility of the dreibein.
} 
Therefore we can provide a quantum theory of the gravity by quantizing 
a classical phase space of the Chern-Simons theory.
In particular, the gravity with a positive cosmological constant 
can be described by taking SL(2;$\mathbb{C}$) as the gauge group.

The Einstein-Hilbert action is written by using the Chern-Simons action 
as follows,
\begin{eqnarray}
\label{Einstein}
S_{EH} &=& 
   \frac{ik}{4\pi}\int_M \mbox{Tr}\left(AdA+\frac23 A^3 \right)
    -\frac{ik}{4\pi}\int_M \mbox{Tr}\left(\bar{A}d\bar{A}
       +\frac23 \bar{A}^3 \right) \nonumber \\
  &=& -\frac{k}{\pi l}\int_M \mbox{Tr}\left[e\left(d\omega+\omega^2\right)
            -\frac{1}{3l^2}e^3\right]
               +\frac{k}{2\pi l}\int_{\partial M} \mbox{Tr} \ e \omega 
       \nonumber \\
  &=& \frac{1}{16\pi G}\int_M d^3x \sqrt{-g}\left(R-\frac{2}{l^2}\right)
       +\frac{1}{8\pi G}\int_{\partial M} \mbox{Tr} \ e \omega,
\end{eqnarray}
where $k=\frac{l}{4G}$ and we decomposed SL(2;$\mathbb{C}$) connections 
$A, \ \bar{A}$ to dreibein $e^a$ and spin connection $\omega^a$,
\begin{equation}
\label{A-e-w}
A=\frac{i}{l}e + \omega, \hspace{10mm}
\bar{A}=-\frac{i}{l}e + \omega.
\end{equation}
The three-dimensional space $M$ is parametrized by 
$\tau, \ z,$ and $\bar{z}$.
In this paper we consider a two-dimensional boundary of $M$ 
at $\tau \longrightarrow -\infty$ which is parametrized 
by $z$ and $\bar{z}$.
In the above form (\ref{Einstein}), the gravitational action 
has a boundary contribution.
We note the fact that this boundary term has a well-known form 
for solutions of the equations of motion,
\begin{equation}
\label{b-term}
 \frac{1}{8\pi G}\int_{\partial M} \mbox{Tr} \ e \omega 
  = -\frac{i}{8\pi G}\int_{\partial M}dz\wedge d\bar{z}
      \frac{\sqrt{h}}{l},
\end{equation}
where $h$ is the determinant of an induced metric on $\partial M$.
The equations of motion of the Chern-Simons theory are the flat
connection conditions which are interpreted as the torsion-free
condition and the Einstein equation with a positive cosmological constant 
$\Lambda=1/l^2$ in terms of dreibein and spin connection.
The flat SL(2;$\mathbb{C}$) connections can be solved as  
\begin{equation}
\label{A-G}
A=G^{-1}dG, \hspace{10mm} 
\bar{A}=\bar{G}^{-1}d\bar{G}.
\end{equation}
For the equations (\ref{A-e-w}) to hold,
$G$ and $\bar{G}$ must satisfy the following relation 
\begin{equation}
\label{G-G}
\bar{G}=\epsilon^{-1} \left(G^{-1}\right)^\dagger \epsilon,
\end{equation}
where the relation 
$\epsilon^{-1} J^a \epsilon=-J_a=-\left(J^a\right)^\dagger$
is useful.
That is, only if this condition is satisfied, the SL(2;$\mathbb{C}$) 
Chern-Simons theory can be regarded as a gravity.

The correspondence between SL(2;$\mathbb{C}$) flat connections and 
the Kerr-de Sitter geometries can be made by the following ways:
first we define a mapping, 
SL(2;$\mathbb{C}$)$\times$SL(2;$\mathbb{C}$)
    $\longrightarrow$SL(2;$\mathbb{C}$)
by
\begin{equation}
\left(G, \ \bar{G}\right)  \longmapsto {\cal G}=G\bar{G}^{-1},
\end{equation}
where it is assumed that $G$ and $\bar{G}$ satisfy (\ref{G-G}).
Thus ${\cal G}$ satisfies 
\begin{equation}
\label{eGe}
\bar{{\cal G}}={\cal G}^{-1} 
 \hspace{5mm} \mbox{or equivalently} \hspace{5mm} 
   {\cal G}=\epsilon {\cal G}^\dagger\epsilon^{-1}.
\end{equation}
${\cal G}$ is invariant under the right diagonal SU(1,1) transformation 
which preserves the relation (\ref{G-G}), 
\begin{equation}
\label{pairing}
G \longmapsto Gg, \hspace{5mm} \bar{G} \longmapsto \bar{G}g, 
\end{equation}
Since ${\cal G}$ satisfies (\ref{eGe}), we can identify ${\cal G}$ with 
${\cal G}_0$ (\ref{cal-G_0}) corresponding to the Kerr-de Sitter space-time.
We consider the following decomposition of ${\cal G}_0$, 
\begin{eqnarray}
{\cal G}_0 &=& G_0\bar{G}_0^{-1}, \nonumber \\
G_0 &=& \left(\begin{array}{cc}
        \frac{a+1}{2\sqrt{a}}z^{\frac{a-1}{2}} & 
          -\frac{i}{\sqrt{a}}z^{\frac{a+1}{2}} \\
        -\frac{i(a-1)}{2\sqrt{a}}z^{-\frac{a+1}{2}} & 
          \frac{1}{\sqrt{a}}z^{-\frac{a-1}{2}}
       \end{array}\right)
      \left(\begin{array}{cc}
        e^{\frac{1}{2}\tau} &  0 \\
      0 & e^{-\frac{1}{2}\tau}
      \end{array}\right), \\
\bar{G}_0 &=&  \left(\begin{array}{cc}
     \frac{1}{\sqrt{\bar{a}}}\bar{z}^{-\frac{\bar{a}-1}{2}} & 
       \frac{i(\bar{a}-1)}{2\sqrt{\bar{a}}}\bar{z}^{-\frac{\bar{a}+1}{2}} \\
     \frac{i}{\sqrt{\bar{a}}}\bar{z}^{\frac{\bar{a}+1}{2}} & 
       \frac{\bar{a}+1}{2\sqrt{\bar{a}}}\bar{z}^{\frac{\bar{a}-1}{2}}
       \end{array}\right)
 \left(\begin{array}{cc}
  e^{-\frac{1}{2}\tau} &  0 \\
  0 & e^{\frac{1}{2}\tau}
	    \end{array}\right)
\end{eqnarray}
Then we construct the flat connections (\ref{A-G}) from $G_0$ and $\bar{G}_0$,
\begin{eqnarray}
\label{A_0}
A_0 &=& G_0^{-1}dG_0 =
  \left(\begin{array}{cc}
   \frac{1}{2}d\tau & -ie^{-\tau}dz \\
   -\frac{ib_0}{c/6}e^{\tau}\frac{dz}{z^2} & -\frac{1}{2}d\tau 
       \end{array}\right), \nonumber \\
\bar{A}_0 &=& \bar{G}_0^{-1}d\bar{G}_0 =
  \left(\begin{array}{cc}
   -\frac{1}{2}d\tau & 
      \frac{i\bar{b}_0}{c/6}e^{\tau}\frac{d\bar{z}}{\bar{z}^2} \\
   ie^{-\tau}d\bar{z} & \frac{1}{2}d\tau 
       \end{array}\right).
\end{eqnarray}
The dreibein obtained from $A_0, \ \bar{A}_0$ are written in the form 
$e=\frac{l}{2i}\left(A_0-\bar{A}_0\right)
  =\frac{l}{2i}\bar{G}_0^{-1}\left({\cal G}_0^{-1}d{\cal G}_0\right)\bar{G}_0$.
Thus these SL(2;$\mathbb{C}$) flat connections provide descriptions 
of the Kerr-de Sitter geometry in the Chern-Simons gravity, that is,  
$ds^2=2 \mbox{Tr} e^2=ds^2_{(b_0, \bar{b}_0)}$.

\subsection{Phase Space of de Sitter Gravity}
In studies of the three-dimensional anti-de Sitter gravity, 
a generalization of anti-de Sitter and the BTZ black hole solutions 
was considered~\cite{CHD-BO, Banados-Ortiz, Banados1, NUY}.
Applying their methods to the de Sitter case, 
we find the following solutions,
\begin{equation}
\label{general-solution}
 A_b = 
  \left(\begin{array}{cc}
   \frac{1}{2}d\tau & -ie^{-\tau}dz \\
   -\frac{ib(z)}{c/6}e^{\tau}dz & -\frac{1}{2}d\tau 
       \end{array}\right),
\hspace{10mm}
\bar{A}_{\bar{b}} = 
  \left(\begin{array}{cc}
   -\frac{1}{2}d\tau & 
      \frac{i\bar{b}(\bar{z})}{c/6}e^{\tau}d\bar{z} \\
   ie^{-\tau}d\bar{z} & \frac{1}{2}d\tau 
       \end{array}\right).
\end{equation}
Here $b(z)$ and $\bar{b}(\bar{z})$ have to be invariant under 
$z\rightarrow e^{2\pi i}z$ to be consistent with the coordinate
transformation (\ref{coord-trans}).
These SL(2;$\mathbb{C}$) flat connections are obtained as solutions of 
the equations of motion under the gauge fixing condition 
$A_\tau=-iJ^0, \ \bar{A}_\tau=iJ^0$ and the following conditions 
which are consistent with the Strominger's boundary conditions to be 
asymptotically de Sitter \cite{Strominger},
\begin{eqnarray}
\label{cond1}
A_{\bar{z}} &\longrightarrow & 0, \hspace{10mm}
   \bar{A}_z \longrightarrow 0,  \\
\label{cond2}
A_z &\hspace{-3mm} \longrightarrow \hspace{-3mm}& 
  \left(\begin{array}{cc}
   0 & -ie^{-\tau} \\
   {\cal O}\left(e^{\tau}\right) & 0
	\end{array}\right), \hspace{5mm}
\bar{A}_{\bar{z}} \longrightarrow 
   \left(\begin{array}{cc}
   0 & {\cal O}\left(e^{\tau}\right) \\
   ie^{-\tau} & 0 
         \end{array}\right), 
 \hspace{5mm}
   \mbox{at} \ \tau \rightarrow -\infty.
\end{eqnarray}

The degrees of freedom of the classical solutions are two functions 
$b(z)$ and $\bar{b}(\bar{z})$, and thus the phase space 
of the three-dimensional de Sitter gravity is composed of these
functions.
The asymptotic Virasoro algebras which act on this phase space are 
realized as residual gauge transformation, 
\begin{equation}
\eta_f=\left(\begin{array}{cc}
	\frac{1}{2}f' & -ife^{-\tau} \\
         -i\left(\frac{1}{2}f''+\frac{b}{c/6}f\right)e^{\tau} & 
           -\frac{1}{2}f
	      \end{array}\right), 
 \hspace{10mm}
\bar{\eta}_{\bar{f}}=\left(\begin{array}{cc}
    -\frac{1}{2}\bar{f}' & 
      i\left(\frac{1}{2}\bar{f}''+\frac{\bar{b}}{c/6}\bar{f}\right)e^{\tau} \\
    i\bar{f}e^{-\tau} & 
      \frac{1}{2}\bar{f}
	      \end{array}\right),
\end{equation}
where $f=f(z)$ and $\bar{f}=\bar{f}(\bar{z})$ should be invariant under
$z\rightarrow e^{2\pi i}z$, also.
These transformations preserve the forms of the flat SL(2;$\mathbb{C}$) 
connections (\ref{general-solution}) and generate the infinitesimal 
deformations of $b(z)$ and $\bar{b}(\bar{z})$.
\begin{eqnarray}
\delta_f A=\left(
      \begin{array}{cc}
	 0 & 0 \\
	 -\frac{i\delta_f b}{c/6}e^\tau dz & 0
      \end{array}\right), 
 \hspace{10mm}
\delta_{\bar{f}} \bar{A}=\left(
      \begin{array}{cc}
   	 0 & \frac{i\delta_{\bar{f}}\bar{b}}{c/6} e^\tau d\bar{z} \\
         0 & 0
      \end{array}\right).
\end{eqnarray}
The deformations of $b(z)$ and $\bar{b}(\bar{z})$ are given 
by the coadjoint actions of the Virasoro algebras with the central
charge $c$,
\begin{equation}
\label{coadjoint-action}
\delta_f b=fb'+2f'b+\frac{c}{12}f''', 
 \hspace{10mm}
\delta_{\bar{f}} \bar{b}=\bar{f}\bar{b}'
     +2\bar{f}'\bar{b}+\frac{c}{12}\bar{f}'''.
\end{equation}
The diffeomorphism corresponding to these gauge transformations is  
generated by the following vector fields,
\begin{eqnarray}
 \xi^\mu \partial_\mu=\frac{1}{2}(f'+\bar{f}')\partial_\tau
     +\left\{f+\frac{\frac{\bar{b}}{c/6}e^{2\tau}f''-\bar{f}''}
                {2\left[\frac{b\bar{b}}{(c/6)^2}e^{2\tau}-e^{-2\tau}\right]}
      \right\}\partial_z
     +\left\{\bar{f}+\frac{\frac{b}{c/6}e^{2\tau}\bar{f}''-f''}
                {2\left[\frac{b\bar{b}}{(c/6)^2}e^{2\tau}-e^{-2\tau}\right]}
      \right\}\partial_{\bar{z}}.
\end{eqnarray}
The flat connections (\ref{general-solution}) describe 
a deformation of the Kerr-de Sitter space-time.
Its metric $ds^2_{(b, \bar{b})}$ is 
\begin{equation}
\label{spacetime-metric}
ds^2_{(b, \bar{b})}=l^2\left[- d\tau^2 +\frac{b(z)}{c/6}dz^2
                +\frac{\bar{b}(\bar{z})}{c/6}d\bar{z}^2
                +\left(e^{-2\tau}+\frac{b(z)\bar{b}(\bar{z})}{(c/6)^2}
                  e^{2\tau}\right)dzd\bar{z} \right].
\end{equation}

The deformed flat connections (\ref{general-solution}) can be solved 
in principle as 
\begin{equation}
\label{G_b}
A_b=G_b^{-1}dG_b, \hspace{20mm} 
\bar{A}_{\bar{b}}=\bar{G}_{\bar{b}}^{-1}d\bar{G}_{\bar{b}}.
\end{equation}
$G_b$ and $\bar{G}_{\bar{b}}$ are generally complicated functions of 
$b(z)$ and $\bar{b}(\bar{z})$, respectively.\footnote{
The explicit forms of $G_b$ and $\bar{G}_{\bar{b}}$ will be given 
in Section 4.}
Then one can provide a region in SL(2;$\mathbb{C}$)/SU(1,1) defined by 
a mapping 
${\cal G}_{(b,\bar{b})}(\tau, z, \bar{z})=G_b\bar{G}_{\bar{b}}^{-1}$. 
We will call this region $Z_{(b,\bar{b})}$.
The space-time corresponding to the deformed solutions 
$A_b$ and $\bar{A}_{\bar{b}}$ is realized as  $Z_{(b,\bar{b})}$  
with the induced metric 
$ds^2_{(b, \bar{b})}=-\frac{l^2}{2}\mbox{Tr}
\left({\cal G}_{(b,\bar{b})}^{-1}d{\cal G}_{(b,\bar{b})}\right)^2$.
The Virasoro algebras with the central charge $c$ act on the phase space
consisting of $b(z)$ and $\bar{b}(\bar{z})$ by coadjoint actions
(\ref{coadjoint-action}). 
Let us consider the coadjoint orbits of $b=\frac{b_0}{z^2}$ 
and $\bar{b}=\frac{\bar{b}_0}{\bar{z}^2}$.
We denote them as $W_{b_0}$ and $\bar{W}_{\bar{b}_0}$ respectively. 
Then $W_{b_0}\times \bar{W}_{\bar{b}_0}$ can be identified with 
the family of $Z_{(b, \bar{b})}$ in which $b(z)$ and $\bar{b}(\bar{z})$ 
lie on $W_{b_0}$ and $\bar{W}_{\bar{b}_0}$ respectively.
Namely, $Z_{\scriptscriptstyle\left(b_0/z^2, \bar{b}_0/\bar{z}^2\right)}$ 
is deformed to $Z_{(b, \bar{b})}$ by coadjoint action of 
the Virasoro algebras.
And the metric $ds^2_{(b_0, \bar{b}_0)}$ is deformed to 
$ds^2_{(b, \bar{b})}$, (\ref{spacetime-metric}).

We again note that each geometry is characterized by its region in 
SL(2;$\mathbb{C}$)/SU(1,1), although all geometries are locally 
equivalent to SL(2;$\mathbb{C}$)/SU(1,1). 
Therefore the degrees of freedom of this system can be regarded as 
ways of taking a region in SL(2;$\mathbb{C}$)/SU(1,1), in other words, 
ways of drawing a two-dimensional surface, which is a boundary of 
a region, in SL(2;$\mathbb{C}$)/SU(1,1).
Thus this degrees of freedom may be interpreted in terms of
a two-dimensional theory.

\section{Quantization of de Sitter Gravity}
In the previous section, we clarified that the phase space  
of the three-dimensional de Sitter gravity consists of the holomorphic 
and anti-holomorphic functions, 
$b(z)$ and $\bar{b}(\bar{z})$ respectively, with invariance 
under $z\rightarrow e^{2\pi i}z$.
Here we first consider an introduction of a symplectic structure 
into the phase space.
It was shown that the Virasoro algebras with {\it pure imaginary} 
central charges are obtained by applying the symplectic reduction method 
to the SL(2;$\mathbb{C}$) Chern-Simons theory~\cite{Park}.\footnote{
In appendix B, we show that the Virasoro algebra with a pure imaginary central
charge is obtained by a standard canonical quantization of 
the SL(2;$\mathbb{C}$)
Chern-Simons theory.}
We here show that there exists a symplectic form which leads to 
the Virasoro algebras with {\it real} central charges after
quantization.
The similar methods to those used in Ref.\cite{NUY} for the asymptotically 
anti-de Sitter space-time, 
e.g., geometric quantization method \cite{Segal, B-R, Witten3, Woodhouse}, 
are applied to the case of de Sitter gravity.
There is an essential difference between these two cases.
While the group which acts on the phase space is $\widehat{Dif\! f S^1}$ 
in anti-de Sitter case, we now must deal with 
the conformal group which is not globally well-defined.
Thus we concentrate on perturbative analyses around the de Sitter and 
Kerr-de Sitter geometries.
Then we consider the unitary representations of the Virasoro algebra 
and discuss the perturbative stability of vacua. 

\subsection{Coadjoint Orbit}
A general element of the Virasoro algebra is written as 
$f(z)\partial_z +ac$ where $c$ is a central element 
and $a$ is a complex number. We denote this as $\left(f, \ a\right)$. 
The commutation relation is given by 
\begin{equation}
\left[(f, \ a_1), \ (g, \ a_2)\right]=
  \left(fg'-f'g, \ \frac{i}{24}\oint \frac{dz}{2\pi i}
      \left(f'''g-fg'''\right)\right).
\end{equation}
Under this commutation relation, 
adjoint vectors $l_m=\left(iz^{m+1}, \ 0\right)$ satisfy 
\begin{equation}
i\left[l_m, \ l_n\right]=(m-n)l_{m+n}+\frac{c}{12}(m^3-m)\delta_{m+n,0}.
\end{equation}
A general coadjoint vector has the form $b(z)dz^2+tc^*$ 
where $t$ is a complex number and $c^*$ a dual element of $c$.
We abbreviate it as $\left(b, \ t\right)$.
The dual pairing between coadjoint and adjoint vectors is defined as 
\begin{equation}
 \langle (b,\ t), \ (g, \ a) \rangle =-i\oint\frac{dz}{2\pi i}b(z)g(z)+ta.
\end{equation}
The coadjoint action of the Virasoro group is given by 
\begin{equation}
\label{delta-b}
 \delta_f (b, \ t)=\left(fb'+2f'b+\frac{t}{12}f''', \ 0\right).
\end{equation}
The dual pairing is invariant under the action of the Virasoro algebra, 
$\displaystyle{\delta_f \langle (b,\ t), \ (g, \ a) \rangle=0}$.
A coadjoint orbit is defined as an orbit of a coadjoint vector by 
coadjoint action of the conformal group.
Let us consider an orbit of $\left(\frac{b_0}{z^2}, \ t\right)$, 
we denote it as $W_{b_0}$, where $b_0$ is a complex number.
From (\ref{delta-b}) for $b=\frac{b_0}{z^2}$ and 
$f=\sum_{n}f_n z^{n+1}$, we find 
\begin{equation}
 \delta_f b=2\sum_n n f_n\left[b_0+\frac{t}{24}(n^2-1)\right] z^{n-2},
\end{equation}
thus $b_0$ is invariant under coadjoint actions of the Virasoro algebra.
There are two classes of orbits.
For general $b_0$, $\delta_f b$ vanishes for $f=z$, thus the stability
group of this type of orbits is dilatation. 
If $b_0=-\frac{t}{24}\left(m^2-1\right)$ where $m \in Z_{> 0}$, 
there are three solutions $f=z, \ z^{1\pm m}$ of $\delta_f b=0$.

For $z_s=s(z)$, the integrated form of (\ref{delta-b}) is given by 
\begin{equation}
\label{b^s}
b^s(z_s)=\left(\frac{dz_s}{dz}\right)^{-2}
  \left[\frac{b_0}{z^2}-\frac{t}{12}\left\{z_s, \ z\right\}\right],
\end{equation}
where $\left\{z_s, \ z\right\}$ is the Schwarzian derivative 
$\left\{z_s, \ z\right\}\equiv\frac{d^3z_s/dz^3}{dz_s/dz}
 -\frac{3}{2}\left(\frac{d^2z_s/dz^2}{dz_s/dz}\right)^2$.
Then $b^s$ is an element of $W_{b_0}$.
The coadjoint orbit representations of generators of the Virasoro algebra
$l_m=\left(iz^{m+1}, \ 0\right)$ are given by 
\begin{equation}
\label{rep-lm}
L_m = \langle (b^s, \ t), \ (iz^{m+1}, \ 0) \rangle 
 = \oint\frac{dz_s}{2\pi i} b^s(z_s)z_s^{m+1}.
\end{equation}
Since the central element corresponds to $(0, \ 1)$, 
it has a constant value $c=t$ on the orbit.
Therefore we denote $t$ as $c$.
A canonical symplectic structure is introduced on each coadjoint orbit
by the dual pairing \cite{Witten3, Woodhouse},
\begin{eqnarray}
\Omega|_{b^s}(u, \ v) 
   &=& \langle (b^s, \ t), \ [(u, \ 0), \ (v, \ 0)] \rangle \nonumber \\
   &=& -i\oint\frac{dz_s}{2\pi i}b^s(z_s)\left(uv'-u'v\right)(z_s)
        -\frac{ic}{24}\oint\frac{dz_s}{2\pi i}\left(uv'''-u'''v\right)(z_s),
\end{eqnarray}
where $u=u(z_s)$ and $v=v(z_s)$ are tangent vectors at $b^s$.
It should be noted that the semi-classical analysis is justified 
in the region of large $c$ because the symplectic form $\Omega$ is large
in that region. 
Under this symplectic structure $L_m$'s (\ref{rep-lm}) satisfy 
the classical Virasoro algebra
\begin{equation}
\label{cl-Virasoro}
i\left\{L_m, \ L_n\right\}
   =(m-n)L_{m+n}+\frac{c}{12}(m^3-m)\delta_{m+n,0}.
\end{equation}

\subsection{Quantization of Coadjoint Orbits}
We first make a perturbative analysis of the orbit $W_{b_0}$ 
at the classical level.
We set 
\begin{equation}
z_s=s(z)=z+i\sum_{n\neq 0}s_n z^{-n+1},
\end{equation}
where $s_n$ is a complex number. 
Then $b^s$ is expanded as the following form from the formula (\ref{b^s})
\begin{eqnarray}
b^s(z_s)&=&\frac{b_0}{z^2}
             +2i\sum_{n}(n-1)\left[b_0+\frac{c}{24}n(n+1)\right]
              s_n z^{-n-2} \nonumber \\
        & & \hspace*{-15mm} -3\sum_{m,n}(m-1)(n-1)
            \left[b_0+\frac{c}{24}(m^2+n^2+mn+m+n)\right]s_m s_n z^{-m-n-2}
             +{\cal O}(s^3).
\end{eqnarray}
The explicit form of the symplectic form is 
\begin{equation}
\Omega(s, \ s')=-i\sum_m\left[2mb_0+\frac{c}{12}(m^3-m)\right]s_m s'_{-m} 
    +{\cal O}(s^3).
\end{equation}
From the inverse matrix of $\Omega$, the following Poisson bracket is 
obtained 
\begin{equation}
\label{s-s}
\left\{s_m, \ s_n\right\} = -i\left[2mb_0+\frac{c}{12}(m^3-m)\right]^{-1}
         \delta_{m+n,0}+{\cal O}(s).
\end{equation}
The expressions for the generators of the Virasoro algebra are 
\begin{eqnarray}
\label{exp-lm}
L_m &=& 2im\left[b_0+\frac{c}{24}(m^2-1)\right]s_{m}+{\cal O}(s^2),
        \qquad {\rm for} \ m\neq 0, \\
\label{exp-l0}
L_0&=&b_0+\sum_{n}n^2 \left[b_0+\frac{c}{24}(n^2-1)\right]s_{-n}s_n
      +{\cal O}(s^3).
\end{eqnarray}
A discussion of the stability of vacua for ``energy function'' $L_{0}$ 
in the anti-de Sitter gravity was
provided by the similar perturbative method \cite{NUY}.  
In that case, the reality condition $(s_n)^*=s_{-n}$ was imposed 
because the coordinate corresponding to $z$ was a real parameter
which parametrizes $S^1$. Therefore $s_n s_{-n}$ was positive definite 
and a condition for vacua to be stable was obtained from the requirement 
that $L_0$ should be bounded from below.
However $z$ is now a complex coordinate and thus no restriction is imposed
on $s_n$ at the classical level. 

We next consider a semi-classical quantization of the orbits.
It is performed by replacing the Poisson bracket (\ref{s-s}) to $-i$ times 
the commutator.
Then we find that $s_n$ are essentially creation and annihilation 
operators up to their normalizations.
We are interested in the unitary representations of 
the Virasoro algebra in which $(L_m)^\dagger=L_{-m}$, 
in particular $(L_0)^\dagger =L_0$, holds.
This unitarity condition leads to the conditions $(s_m)^\dagger=s_{-m}$ 
and $b_0 \in \mathbb{R}$ up to this order of perturbations.
\footnote{This type of Hermitian conjugation is also considered
in view of the dS/CFT correspondence in Ref.\cite{BMS}.} 
The stability of the classical vacua means that $L_0$ should be bounded
below at $b(z)=\frac{b_0}{z^2}$. 
It holds true only when 
\begin{equation}
\label{bound}
b_0 \geq 0.
\end{equation}
Therefore only orbits with $b_0 \in \mathbb{R}_{\geq 0}$ provide 
the unitary representations of the Virasoro algebra 
with the stable vacua.

The quantum Virasoro algebra is given by replacing the Poisson bracket 
to $-i$ times the commutator in the classical Virasoro algebra 
(\ref{cl-Virasoro}),
\begin{equation}
\label{quantum-Virasoro}
\left[L_m, \ L_n\right]
   =(m-n)L_{m+n}+\frac{c}{12}(m^3-m)\delta_{m+n,0}.
\end{equation}
Each unitary representation of the Virasoro algebra is labeled by 
its highest weight state $|h\rangle$.
The highest weight state satisfies 
$L_0|h\rangle=h|h\rangle$ and $L_n|h\rangle=0$ for $n\geq 1$.
In particular $|0\rangle$ satisfy $L_{-1}|0\rangle=0$ in addition to 
the above conditions.
The unitary irreducible representations ${\cal V}_{h=b_0}$ corresponding to 
$W_{b_0} \ (b_0 >0)$ are constructed 
by acting the Virasoro generators $L_{-n}$ ($n\geq 1$) on $|b_0\rangle$.
The unitary irreducible representation ${\cal V'}_{h=0}$ corresponding 
to $W_{b_0=0}$ is given by acting $L_{-n}$ ($n\geq 2$) on $|0\rangle$.

\subsection{Quantization of Gravity in dS$\bf _3$}
The phase space of the three-dimensional de Sitter gravity is composed 
of de Sitter, Kerr-de Sitter geometries and their Virasoro 
deformations.
The de Sitter space and its Virasoro deformations are identified with 
the product of the Virasoro coadjoint orbits 
$W_{b_0=0}\times\bar{W}_{\bar{b}_0=0}$. 
The quantizations of the orbits lead to the unitary irreducible
representations of the Virasoro algebras 
${\cal V'}_{h=0}\times \bar{{\cal V'}}_{\bar{h}=0}$.
The Kerr-de Sitter geometry with $b=\frac{b_0}{z^2}$ and
$\bar{b}=\frac{\bar{b}_0}{\bar{z}^2}$ and their deformations 
$Z_{(b, \bar{b})}$  are identified with the product of the orbits 
$W_{b_0}\times\bar{W}_{\bar{b}_0}$.
If $b_0$ is real, that is $b_0=\bar{b}_0$, and $b_0>0$, 
the quantizations of the orbits lead to the unitary irreducible
representations of the Virasoro algebras 
${\cal V}_{b_0}\times \bar{{\cal V}}_{b_0}$.
The unitarity condition that $b_0$ is real means $J=0$, 
thus $b_0=\frac{c}{24}(1-\mu)$ from (\ref{b_0}).
In addition to that, the stability condition requires $b_0 \geq 0$. 
Therefore $\mu$ is restricted in the region $\mu \leq 1$.
Geometries with excess angles ($\mu >1$) are forbidden 
by these conditions.
As a result, the quantizations of deformations only of de Sitter
and conical geometries without angular momenta provide the unitary theories
with stable vacua. 
On the other hand, quantizations of the Kerr-de Sitter space-times 
with non-vanishing angular momenta and their deformations 
do not give unitary theories.

We provide some comments for the geometries with $J=0$ and $\mu<0$.
As noted in the previous section, 
these geometries seem to be unphysical in the Schwarzschild  
coordinates. However, according to the discussion in this section, 
they are naturally included in the phase space 
constructed here and are represented by the following region in 
SL(2;$\mathbb{C}$)/SU(1,1), 
\begin{eqnarray}
 {\cal G}_{(\mu<0)} &\equiv& \frac{1}{m}
 \left(\begin{array}{cc}
  -\frac{i(1+i\sqrt{m})}{2}z^{-\frac{1-i\sqrt{m}}{2}} & 
    -z^{\frac{1+i\sqrt{m}}{2}} \\
  \frac{1-i\sqrt{m}}{2}z^{-\frac{1-i\sqrt{m}}{2}} & 
    -iz^{\frac{1-i\sqrt{m}}{2}}
       \end{array}\right)
 \left(\begin{array}{cc}
  e^{\tau} &  0 \\
  0 & e^{-\tau}
       \end{array}\right) \nonumber \\
&& \hspace*{20mm} \times 
  \left(\begin{array}{cc}
    \frac{i(1-i\sqrt{m})}{2}\bar{z}^{-\frac{1+i\sqrt{m}}{2}} & 
      -\frac{1+i\sqrt{m}}{2}\bar{z}^{-\frac{1-i\sqrt{m}}{2}} \\
    \bar{z}^{\frac{1-i\sqrt{m}}{2}} & 
      i\bar{z}^{\frac{1+i\sqrt{m}}{2}}
       \end{array}\right),
\end{eqnarray}
with $m\equiv-\mu$.
The induced metric on this region is given by 
\begin{equation}
ds^2_{(\mu<0)}=l^2\left[- d\tau^2 +\frac{1+m}{4}\frac{dz^2}{z^2}
                +\frac{1+m}{4}\frac{d\bar{z}^2}{\bar{z}^2}
                +\left(e^{-2\tau}+\frac{(1+m)^2}{16}\frac{1}{z^2 \bar{z}^2}
                  e^{2\tau}\right)dzd\bar{z} \right],
\end{equation}
that is $ds^2_{(\mu<0)}=ds^2_{(b_0, \bar{b}_0)}$ with 
$b_0=\bar{b}_0=\frac{c}{24}(1+m)$.
The Virasoro deformations of these geometries can be defined by the same way 
as those explained until here. 
After quantization, furthermore, the states of these geometries 
are allowed under our consideration about the unitarity and stability
at the perturbative level.

\section{On Three-dimensional de Sitter Gravity and Liouville Field Theory}
It is well-known that the Chern-Simons action on a manifold with
boundaries reduces to the chiral WZNW actions 
on the boundaries \cite{Elitzur}.
It was shown that the action of the three-dimensional anti-de Sitter 
gravity which is written by two SL(2;$\mathbb{R}$) Chern-Simons actions  
reduces to the non-chiral WZNW action and furthermore 
it reduces to the Liouville field theory under the boundary conditions 
being the asymptotically anti-de Sitter space-time \cite{CHD-BO}.
Recently analogous analyses were done in \cite{Klemm1}
for the three-dimensional de Sitter gravity with the asymptotically past
de Sitter space-time.

\subsection{Reduction to Liouville Field Theory}
We briefly review processes of the reduction to the Liouville field
theory \cite{Klemm1} in order to fix the notations we use here.

The Einstein action is given by the SL(2;$\mathbb{C}$) Chern-Simons 
action (\ref{Einstein}).
The reduction is carried out by two stages.
At the first stage, we impose the following reduction conditions,
\begin{equation}
A_{\bar{z}}\longrightarrow 0, \hspace{20mm}
\bar{A}_z\longrightarrow 0,
\end{equation}
at the infinitely past $\tau \longrightarrow -\infty$.
These conditions are same as (\ref{cond1}). 
Under these conditions and the gauge fixing 
$A_\tau=-iJ^0, \ \bar{A}_\tau=iJ^0$, the flat connections are written as
\begin{equation}
A=G^{-1}dG, \hspace{20mm}
\bar{A}=\bar{G}^{-1}d\bar{G},
\end{equation}
where $G$ and $\bar{G}$ are group elements of SL(2;$\mathbb{C}$) 
and behave at $\tau \longrightarrow -\infty$ as
\begin{eqnarray}
G \longrightarrow g(z, \ \bar{z})
   \left(\begin{array}{cc}
    e^{\frac{1}{2}\tau} & 0 \\
    0 & e^{-\frac{1}{2}\tau}
	 \end{array}\right),  
  \hspace{10mm}
\bar{G} \longrightarrow \bar{g}(z, \ \bar{z})
   \left(\begin{array}{cc}
    e^{-\frac{1}{2}\tau} & 0 \\
    0 & e^{\frac{1}{2}\tau}
	 \end{array}\right).
\end{eqnarray}
Substituting these flat connections into the SL(2;$\mathbb{C}$) 
Chern-Simons action, we obtain two chiral WZNW action in terms of 
$g(z, \ \bar{z})$ and $\bar{g}(z, \ \bar{z})$ respectively.
Then changing the variables,
\begin{equation}
\label{H-G}
H=G^{-1}\bar{G}, \hspace{10mm}
h=g^{-1}\bar{g},
\end{equation}
two chiral WZNW actions are combined to the non-chiral WZNW action,
\begin{equation}
\label{WZNW}
S=\frac{ik}{4\pi}\int_{\partial M}dz\wedge d\bar{z} \ 
  \mbox{Tr} \left(h^{-1}\partial_zh h^{-1}\partial_{\bar{z}}h\right)
   +\frac{ik}{12\pi}\int_M \mbox{Tr} \left(H^{-1}dH\right)^3.
\end{equation}
Here we provide some comments to this WZNW action.
$H$ and $h$ must satisfy $H=\epsilon H^\dagger \epsilon^{-1}$ and 
$h=\epsilon h^\dagger \epsilon^{-1}$
for this theory to describe the de Sitter gravity.
Thus this is actually SL(2;$\mathbb{C}$)/SU(1,1) WZNW theory.
Next the composition (\ref{H-G}) is different from the pairing 
(\ref{pairing}) which we defined in the previous section to clarify 
the connection between the SL(2;$\mathbb{C}$) flat connections and
the space-time metric.

At the second stage of the reduction, conditions
which are on-shell equivalent to the equations (\ref{cond2}) 
in the previous section are imposed on the conserved currents 
of the WZNW theory. 
We can explicitly describe them by using the Gauss
decomposition
\begin{equation}
\label{Gauss-dec}
h=\left(\begin{array}{cc}
         1 & 0 \\
         X & 1	
        \end{array}\right)
  \left(\begin{array}{cc}
         e^{-\frac{1}{2}\tilde{\Phi}} &  0 \\
         0 & e^{\frac{1}{2}\tilde{\Phi}}
        \end{array}\right)
  \left(\begin{array}{cc}
         1 & Y \\
         0 & 1 
        \end{array}\right).
\end{equation}
Then the constraints are written by 
\begin{equation}
\label{W2L}
 \left\{\begin{array}{l}
  J^0_z=
   2i\left(-\frac{1}{2}\partial_z \tilde{\Phi} 
      -e^{-\tilde{\Phi}}X\partial_z Y\right)
    =0 \\
  \frac{1}{2}\left(J^1_z+iJ^2_z\right)
   =-e^{-\tilde{\Phi}}\partial_z Y
    =-i,
	\end{array}\right. 
\hspace{10mm}
 \left\{\begin{array}{l}
  \bar{J}^0_{\bar{z}}
   =2i\left(-\frac{1}{2}\partial_{\bar{z}} \tilde{\Phi} 
     -e^{-\tilde{\Phi}}Y\partial_{\bar{z}} X\right)
    =0 \\
  \frac{1}{2}\left(\bar{J}^1_{\bar{z}}-i\bar{J}^2_{\bar{z}}\right)
   =-e^{-\tilde{\Phi}}\partial_{\bar{z}} X
     =-i,
	\end{array}\right.
\end{equation}
where $J_z=\partial_z h h^{-1}$ and $\bar{J}_{\bar{z}}
=h^{-1}\partial_{\bar{z}} h $.
After applying the constraints, the WZNW action (\ref{WZNW}) reduces to 
one of the Liouville field theory,
\begin{equation}
S=\frac{ik}{4\pi}\int_{\partial M} \ dz\wedge d\bar{z}
  \left(\frac{1}{2}\partial_z\tilde{\Phi} \partial_{\bar{z}}\tilde{\Phi}
   +2e^{\tilde{\Phi}}\right).
\end{equation}
Moreover transforming to the conformal gauge of the metric on 
$\partial M$ and redefining $\tilde{\Phi}=\gamma\Phi$ , 
the action is written as  
\begin{equation}
S=\frac{1}{4\pi}\int_{\partial M} d^2x 
 \left[\frac{1}{2}h^{ij}\partial_i \Phi\partial_j \Phi
  +\frac{\lambda}{2\gamma^2}e^{\gamma\Phi}
   +\frac{Q}{2}R\Phi\right],
\end{equation}
where $Q=\frac{2}{\gamma}=\sqrt{\frac{l}{2G}}$ and 
$\lambda=\frac{16}{l^2}$.

\subsection{3d gravity and Liouville Field Theory}
General solutions of the equation of motion of the Liouville theory 
are locally given by the following form \cite{Seiberg}
\begin{equation}
\label{L-sol}
e^{\gamma\Phi}=\frac{\partial_z F \partial_{\bar{z}}\bar{F}}
                             {(1-F\bar{F})^2}
\end{equation}
where $F$ is a holomorphic function $F=F(z)$ and 
$\bar{F}$ is anti-holomorphic $\bar{F}=\bar{F}(\bar{z})$.
While $e^{\gamma\Phi}$ must be invariant 
under $z\rightarrow e^{2\pi i}z$, $F$ and $\bar{F}$ are not required 
to have this invariance.

We give explicit connections between these classical solutions and
three-dimensional geometries.
From the constraints (\ref{W2L}) we find 
\begin{equation}
X=\frac{i\gamma}{2}\partial_z \Phi, \hspace{10mm}
Y=\frac{i\gamma}{2}\partial_{\bar{z}} \Phi.
\end{equation}
Substituting these relations and the classical solutions (\ref{L-sol}) 
into the Gauss decomposition of $h$ (\ref{Gauss-dec}), 
$h$ can be written as the decomposed form into holomorphic and 
anti-holomorphic parts,
\begin{equation}
h\equiv g(z)^{-1}\bar{g}(\bar{z})
 = \left(\begin{array}{cc}
    \frac{1}{\sqrt{\partial_zF}} & \frac{iF}{\sqrt{\partial_z F}} \\
    \frac{i}{2}\frac{\partial_z^2 F}
                   {\left(\partial_z F\right)^{3/2}} &
     \frac{\left(\partial_z F\right)^2-\frac{1}{2}F\partial_z^2 F}
          {\left(\partial_z F\right)^{3/2}}
	\end{array}\right)
  \left(\begin{array}{cc}
   \frac{1}{\sqrt{\partial_{\bar{z}} \bar{F}}} & 
             \frac{i}{2}\frac{\partial_{\bar{z}}^2 \bar{F}}
               {\left(\partial_{\bar{z}} \bar{F}\right)^{3/2}} \\
   \frac{i\bar{F}}{\sqrt{\partial_{\bar{z}} \bar{F}}} &
     \frac{\left(\partial_{\bar{z}} \bar{F}\right)^2
            -\frac{1}{2}\bar{F}\partial_{\bar{z}}^2 \bar{F}}
          {\left(\partial_{\bar{z}} \bar{F}\right)^{3/2}}
	\end{array}\right).
\end{equation}
By following the inverse procedures of the reduction from the Chern-Simon 
theory to the Liouville theory, the SL(2;$\mathbb{C}$) 
flat connections corresponding to (\ref{L-sol}) can be obtained as 
\begin{equation}
A=\left(\begin{array}{cc}
   \frac{1}{2}d\tau & -ie^{-\tau}dz \\
   -\frac{i}{2}\left\{F, \ z\right\}dz & -\frac{1}{2}d\tau
	\end{array}\right), 
  \hspace{5mm}
\bar{A}=\left(\begin{array}{cc}
	 -\frac{1}{2}d\tau & 
            \frac{i}{2}\left\{\bar{F}, \ \bar{z}\right\}d\bar{z} \\
         ie^{-\tau}d\bar{z} & \frac{1}{2}d\tau
	      \end{array}\right).
\end{equation}
We identify $A$ and $\bar{A}$ with the generalized solutions 
$A_b$ and $\bar{A}_{\bar{b}}$ (\ref{general-solution}).
Then $b$ ($\bar{b}$) is written by the Schwarzian derivative of 
$F$ ($\bar{F}$),
\begin{equation}
\label{stress}
b(z)=\frac{c}{12}\left\{F(z), \ z\right\}, \hspace{10mm}
\bar{b}(\bar{z})=\frac{c}{12}\left\{\bar{F}(\bar{z}), \ \bar{z}\right\}.
\end{equation}
That is, $b(z)$ $\left(\bar{b}(\bar{z})\right)$ coincides with 
the stress tensor of the holomorphic (anti-holomorphic) part 
of the Liouville theory.
By using $F$ and $\bar{F}$, $G_b$ and $\bar{G}_{\bar{b}}$ (\ref{G_b}) 
are written as $G_b=g(z)e^{-i\tau J^0}$ and 
$\bar{G}_{\bar{b}}=\bar{g}(\bar{z})e^{i\tau J^0}$.
Thus we can give the explicit form of the region corresponding to 
the deformed space-time,
\begin{eqnarray}
\hspace*{-5mm}
{\cal G}_{(b, \bar{b})}
  \hspace{-2mm}&=& \hspace{-2mm}G_b \bar{G}_{\bar{b}}^{-1} \nonumber \\
  \hspace{-2mm}&=& \hspace{-2mm}
  \left(\begin{array}{cc}
   \frac{\left(\partial_z F\right)^2-\frac{1}{2}F\partial_z^2 F}
          {\left(\partial_z F\right)^{3/2}} & 
   -\frac{iF}{\sqrt{\partial_z F}} \\
   -\frac{i}{2}\frac{\partial_z^2 F}
                   {\left(\partial_z F\right)^{3/2}} &
    \frac{1}{\sqrt{\partial_zF}}
       \end{array}\right)
 \left(\begin{array}{cc}
  e^{\tau} &  0 \\
  0 & e^{-\tau}
       \end{array}\right)
 \left(\begin{array}{cc}
   \frac{\left(\partial_{\bar{z}} \bar{F}\right)^2
            -\frac{1}{2}\bar{F}\partial_{\bar{z}}^2 \bar{F}}
          {\left(\partial_{\bar{z}} \bar{F}\right)^{3/2}} & 
             -\frac{i}{2}\frac{\partial_{\bar{z}}^2 \bar{F}}
               {\left(\partial_{\bar{z}} \bar{F}\right)^{3/2}} \\
   -\frac{i\bar{F}}{\sqrt{\partial_{\bar{z}} \bar{F}}} &
   \frac{1}{\sqrt{\partial_{\bar{z}} \bar{F}}}    
       \end{array}\right),
\end{eqnarray}
Setting $F=F_0\equiv z^a$ and 
$\bar{F}=\bar{F}_0\equiv\bar{z}^{\bar{a}}$, 
the region ${\cal G}_0$ (\ref{cal-G_0}) corresponding to the Kerr-de Sitter
solution is obtained.  

We would like clearly to see the relationship between 
the phase space of the three-dimensional gravity and 
that of the Liouville theory.
Under the conformal transformation $z\longrightarrow z_s=s(z)$, 
$F(z)$ is transformed as scalar, $F^s(z_s)=F(z)$.
Then $b(z)$ (\ref{stress}) correctly transforms as the stress tensor,
\begin{eqnarray}
b(z)\longrightarrow b^s(z_s) 
 &=& \frac{c}{12}\left\{F^s(z_s), \ z_s\right\} \nonumber \\
 &=& \left(\frac{dz_s}{dz}\right)^{-2}
  \left[b(z)-\frac{c}{12}\left\{z_s, \ z\right\}\right].
\end{eqnarray}
The infinitesimal form of the transformation is $\delta_f F=F'f$ 
and $\delta_f b =fb'+2f'b+\frac{c}{12}f'''$, 
where we took $z_s=z-f(z)$ and $f$ is a single valued function on the complex
plane.
Setting $F=F_0$ and $f=-i\sum_n u_n z^{-n+1}$, in particular, 
we consider deformations of the solution 
corresponding to the Kerr-de Sitter geometry,
\begin{equation}
 F_f(z)=F_0+\delta_f F_0=z^a\left(1-ia\sum_n u_nz^{-n}\right).
\end{equation}
For this $F_f$, the stress tensor (\ref{stress}) becomes
\begin{eqnarray}
b_f(z)&=&\frac{1}{z^2}\left\{\frac{c}{24}(1-a^2)
      +\frac{ic}{12}\sum_n n(a^2-n^2)u_{-n}z^n \right. \nonumber \\
&& \hspace*{-13.5mm}   
   \left. -\frac{c}{24}\sum_{m,n}(a+m)(a+n)
      \left[(m+n)a-m^2-n^2-3mn\right]u_{-m}u_{-n}z^{m+n}\right\}
        +{\cal O}(u^3).
\end{eqnarray}
Then we find 
\begin{eqnarray}
L_m &=& \oint\frac{dz}{2\pi i}b(z)z^{m+1}
     = 2im\left[b_0+\frac{c}{24}(m^2-1)\right]u_m +{\cal O}(u^2),
       \hspace{10mm} \mbox{for} \ \ m\neq 0   \nonumber \\
L_0 &=& \oint\frac{dz}{2\pi i}b(z)z
     = b_0 + \sum_n n^2\left[b_0+\frac{c}{24}(n^2-1)\right]u_{-n}u_n
       +{\cal O}(u^3),
\end{eqnarray}
where $b_0=\frac{c}{24}(1-a^2)$.
These coincide with the perturbative expressions, 
(\ref{exp-lm}) and (\ref{exp-l0}), of the Virasoro generators 
if we identify $u_m$ with $s_m$.
Therefore we obtained the correspondences between the infinitesimal 
Virasoro deformations of the Kerr-de Sitter geometries and those of 
configurations of the Liouville field. 
Thus the quantization of this Liouville theory
is expected to reproduce the same unitary irreducible
representations of the Virasoro algebras 
as those of the three-dimensional gravity obtained 
in the previous section. 

In the quantum Liouville field theory \cite{Seiberg, Teschner}, 
the primary field is given by 
$e^{\alpha\Phi} \ (\alpha\in\mathbb{C})$ with weight 
$-\frac{1}{2}\alpha(\alpha-Q)$. 
And the primary state is defined by 
$\displaystyle{|\alpha\rangle \equiv \lim_{z\rightarrow 0}e^{\alpha\Phi}(z)|0\rangle}$ 
which satisfies
\begin{equation}
L_0|\alpha\rangle = 
   -\frac{1}{2}\alpha(\alpha-Q)|\alpha\rangle,
\hspace{10mm}
L_n|\alpha\rangle = 0 \ \ \ \mbox{for} \ n\geq 1,
\end{equation}
where $|0\rangle$ is the SL(2;$\mathbb{C}$) invariant vacuum, 
$L_n|0\rangle =0$ for $n\geq-1$.
The representations of the Virasoro algebra are constructed by acting 
$L_{-n} (n\geq 1)$ on $|\alpha\rangle$ and $L_{-n} (n\geq 2)$ on
$|0\rangle$, respectively. 

The primary state $|\alpha\rangle\otimes|\alpha\rangle$ is identified 
with the state of the point particle without angular momentum 
in the 3d gravity by matching the conformal weight with 
$b_0=\frac{c}{24}(1-\mu)$, 
\begin{equation}
 \label{alpha}
 \alpha=\sqrt{\frac{c}{12}}\left[ \ 1-\sqrt{1-\frac{b_0}{c/24}} \ \right], 
\end{equation}
where $0\leq b_0 \leq \frac{c}{24}$, thus $\alpha$ is real 
and $0 \leq \alpha \leq \sqrt{\frac{c}{12}}$. 
This upper bound for the value of $\alpha$ corresponds to the Seiberg
bound. The state with $\alpha$ within this range is non-normalizable.
The excitations on the primary states can be identified with 
the excitations by the deformations of the point particle state, 
as explained in terms of classical perturbations.
In the case of the geometries with $\mu<0$ and $J=0$, 
their counterparts in the Liouville field theory are the primary states 
$|\alpha\rangle\otimes|\alpha\rangle$ with 
$\alpha=\sqrt{c/12}\left[1\pm i\sqrt{m}\right]$ where $m=-\mu>0$.
These states are normalizable because $Re(\alpha)=\sqrt{c/12}$.
These sorts of states are allowed by our discussion about 
the unitarity and stability in the gravity side.
On the other hand, we have concluded that the states of the Kerr-de Sitter 
geometries ($J\neq 0$) and their deformations do not provide 
the unitary theory.
Those states would correspond to states in representations constructed on 
the primary states of the Liouville theory with complex value of $\alpha$ 
which is given by eq.(\ref{alpha}) with 
$b_0=\frac{c}{24}(1-\mu)-\frac{i}{2}J$.
It is known from studies of the Liouville theory 
that these representations with $Re(\alpha)\neq\sqrt{c/12}$ are 
not unitary.

\section{Discussions}
We would like to discuss relations between the holonomies 
of SL(2;$\mathbb{C}$) flat connections and the zero modes 
of the Liouville field.
The geometries we have considered here have the closed paths 
$z\sim e^{2\pi i}z$ and  $\bar{z}\sim e^{-2\pi i}\bar{z}$.
The holonomies of the SL(2;$\mathbb{C}$) flat connections (\ref{A_0}), 
which correspond to the Kerr-de Sitter metric, along these paths 
are given by
\begin{eqnarray}
\mbox{Tr} \ Pe^{\oint A}
 &=&\mbox{Tr} \ G_0^{-1}(z,\tau)G_0(e^{2\pi i}z, \tau)
  =-2\cos(a\pi), \nonumber \\ 
\mbox{Tr} \ Pe^{\oint \bar{A}}
 &=& \mbox{Tr} \ \bar{G}_0^{-1}(\bar{z},\tau)
       \bar{G}_0(e^{-2\pi i}\bar{z}, \tau)
  =-2\cos(\bar{a}\pi), 
\end{eqnarray}
where $a=\sqrt{1-\frac{b_0}{c/24}}$ and 
$\bar{a}=\sqrt{1-\frac{\bar{b}_0}{c/24}}$ as defined before.
Thus the Kerr-de Sitter metric can be characterized by these holonomies 
as well its mass and angular momentum.
Substituting the solution corresponding to this geometry 
in the Liouville theory into the right hand side 
in the eq.(\ref{L-sol}), 
we find 
\begin{equation}
\Phi \sim \frac{1}{\gamma}\ln a + \frac{(a-1)}{\gamma}\ln z
          +\frac{1}{\gamma}\ln \bar{a} 
            + \frac{(\bar{a}-1)}{\gamma}\ln \bar{z}.
\end{equation}
Here we ignore the denominator of the right hand side 
in the eq.(\ref{L-sol}), that is, the effect of the Liouville potential, 
thus this equality is not strict.
However comparing this equation with the following expansion 
$\Phi \sim x-ip\ln z +\mbox{(anti-holomorphic part)}$,  
we obtain 
\begin{equation}
x=\frac{1}{\gamma}\ln a, \hspace{10mm}
p=\frac{i(a-1)}{\gamma}.
\end{equation}
These correspond to the relations which used in the Ref.\cite{NUY} to 
investigate a quantization of three-dimensional gravity 
with a negative cosmological constant in the view of 
the Liouville field theory.
As pointed out in that paper, we should also consider quantization of 
these global degrees of freedom, holonomies, 
within the three-dimensional gravity.

We think that more accurate analyses are needed in order to 
completely understand the quantum nature of the three-dimensional 
de Sitter gravity. 
First we implicitly assumed in this paper that 
the states obtained by quantization of the phase space of the gravity, 
e.g., de Sitter or point particle geometries,  
are normalizable.
On the other hand the primary states corresponding 
to the point particle geometries are non-normalizable
in the Liouville field theory.
To clarify origin of this discrepancy, it is needed 
to study the global structure of the phase space and 
to strictly define the inner product between quantum states 
in the de Sitter gravity.
Next we stated in Section 2 that the flat SL(2;$\mathbb{C}$) 
connections $A$ and $\bar{A}$ can not be independent mutually 
so that the dreibein and spin connection are real quantities.
This requirement connects the holomorphic part of the degrees 
of freedom of the theory with the anti-holomorphic part, that is, 
$\bar{b}(\bar{z})$ should be a complex conjugate function of $b(z)$.
On the other hand, we imposed $L_m^\dagger =L_{-m}$, 
which are relations among the modes of the holomorphic part $b(z)$,  
as the unitarity condition.
It might be possible to adopt other conditions 
which relate $b(z)$ to $\bar{b}(\bar{z})$ 
by Hermitian conjugation.\footnote{
Recently this type of Hermitian conjugations has been discussed 
in Ref.\cite{BBM2}}

In this paper we used the coordinate system corresponding to 
the planar coordinate of de Sitter space.
In general quantum theory might depend on the choice of 
the coordinate system, 
e.g., the vacua corresponding to different coordinates could be 
connected with each other by the Bogoliubov transformations.
Therefore it is meaningful to investigate quantum theory of 
the de Sitter gravity in other coordinate systems, particularly 
in the global coordinate.

\vspace{10mm}
%%% Acknowledgment %%%
\noindent
{\bf \large Acknowledgment} \\
This work of H. U. was supported in part by JSPS Research Fellowships 
for Young Scientists.

\renewcommand{\theequation}{\Alph{section}.\arabic{equation}}
\appendix

\section{Brown-York Stress Tensor}
Here we show that the Brown-York stress tensor \cite{Brown-York}
for general solutions we used coincides with 
$b(z)$ and $\bar{b}(\bar{z})$.

Since the equation (\ref{b-term}) is satisfied for solutions of 
the equations of motion, we consider the action
\footnote{This action originated from the Chern-Simons action is
on-shell equivalent to the gravitational action including the boundary
term and the counter term considered in Ref.\cite{Brown-York,
Strominger, BBM1}.} 
\begin{equation}
S=\frac{1}{16\pi G}\int_M d^3x \sqrt{-g}\left(R-\frac{2}{l^2}\right)
       -\frac{i}{8\pi G}\int_{\partial M}dz\wedge d\bar{z}
        \frac{\sqrt{h}}{l}.
\end{equation}
For the space-time metric
\begin{equation}
ds^2=l^2\left[-dt^2+h_{zz}dz^2+h_{\bar{z}\bar{z}}d\bar{z}^2
   +2h_{z\bar{z}}dzd\bar{z}\right],
\end{equation}
the Brown-York stress tensors are given by 
\begin{eqnarray}
 T_{zz} &=& \frac{2}{\sqrt{h}}\frac{\delta S}{\delta h^{zz}}
   = \frac{1}{8\pi G}\left[R_{zz}-\frac{1}{2}h_{zz}
      \left(R-\frac{2}{l^2}\right)\right]
       +\frac{1}{8\pi Gl}\left[h_{zz}
         +\frac{1}{2}\partial_t h_{zz}\right]_{\partial M}, \\
 T_{\bar{z}\bar{z}} &=& 
  \frac{2}{\sqrt{h}}\frac{\delta S}{\delta h^{\bar{z}\bar{z}}}
   = \frac{1}{8\pi G}\left[R_{\bar{z}\bar{z}}-\frac{1}{2}h_{\bar{z}\bar{z}}
      \left(R-\frac{2}{l^2}\right)\right]
       +\frac{1}{8\pi Gl}\left[h_{\bar{z}\bar{z}}
         +\frac{1}{2}\partial_t h_{\bar{z}\bar{z}}\right]_{\partial M}, \\
T_{z\bar{z}} &=& \frac{1}{\sqrt{h}}\frac{\delta S}{\delta h^{z\bar{z}}}
   = \frac{1}{8\pi G}\left[R_{z\bar{z}}-\frac{1}{2}h_{z\bar{z}}
      \left(R-\frac{2}{l^2}\right)\right]
       +\frac{1}{8\pi Gl}\left[h_{z\bar{z}}
         +\frac{1}{2}\partial_t h_{z\bar{z}}\right]_{\partial M}.
\end{eqnarray}
Substituting the general space-time metric (\ref{spacetime-metric}), 
the bulk terms of the above stress tensors vanish and only boundary 
contributions remain,
\begin{eqnarray}
 T_{zz}=\frac{1}{2\pi l^2}b(z), \hspace{10mm}
 T_{\bar{z}\bar{z}}=\frac{1}{2\pi l^2} \bar{b}(\bar{z}), \hspace{10mm}
 T_{z\bar{z}}=0.
\end{eqnarray}

\section{Canonical Quantization of $SL(2;\mathbb{C})$ Chern-Simons Gravity}
In this appendix, as mentioned in Section 3, we make a canonical
quantization \`a la Dirac of three-dimensional de Sitter gravity 
described by the action (2.13). (The case of three-dimensional 
anti-de Sitter gravity is discussed in Ref.\cite{Banados2, B-B-O, 
Oh, Park2, Banados1}.)    
Then we find the Virasoro algebra
with a {\it pure imaginary} central charge as the asymptotic symmetry.

$\textrm{SL}(2;\mathbb{C})$ Chern-Simons action is defined 
from (\ref{Einstein}):\footnote{$\epsilon^{\tau\rho\sigma}={\rm +}1$.}
\begin{eqnarray}
S_{EH}&=&S_{CS}(A) - S_{CS}(\bar{A}), \nonumber \\
S_{CS}(A)&=&\frac{i k}{4 \pi} \int_{M}\textrm{Tr}\left(A dA +
\frac{2}{3} A^{3}\right)\\
&=&\frac{i k}{4 \pi}\int_{M}\!\!\!\! 
d\tau\wedge d\rho\wedge d\sigma~
\epsilon^{\mu\nu\rho}\!\left(\eta_{ab}A_{\mu}^{a}\partial_{\nu}A_{\rho}^{b}
+\epsilon_{abc}A_{\mu}^{a}A_{\nu}^{b}A_{\rho}^{c}\right)\!, 
\label{eq:CS action}
\end{eqnarray} 
where $A$ is the $sl(2;\mathbb{C})$-valued 1-form related to
dreibein and spin connection by (\ref{A-e-w}).
$M$ is a three-dimensional manifold whose topology is ${\bf R} \times
{\bf R}^{2}$. $M$ is parameterized by $(\tau, \rho, \sigma)$ whose region 
is $(-\infty < \tau < \infty,~0 \le \rho < \infty,~0 \le \sigma < 2
\pi)$. We concentrate on $S_{CS}(A)$ part only in the sequel
since the discussion on $S_{CS}(\bar{A})$ part is parallel.

$M$ is a non-compact manifold with a two-dimensional boundary 
$\tau\rightarrow -\infty$ and $\rho\rightarrow\infty$. 
Thus we require the boundary condition for $A$ 
so that the action (\ref{eq:CS action}) is differentiable with respect
to $A_{\rho}$ and $A_{\sigma}$:
\begin{eqnarray}
A_{\bar{z}}=0 \quad {\rm at} \quad \tau \rightarrow -\infty,
\label{eq:boundary condition}
\end{eqnarray}
where $z=e^{\rho +i\sigma}$ and 
$A_{z}=(A_{\rho}-iA_{\sigma})/2z,~
A_{\bar{z}}=(A_{\rho}+iA_{\sigma})/2\bar{z}$.
This boundary condition is the same as (\ref{cond1}).
As for the boundary $\rho\rightarrow\infty$ we include the 
boundary term in the Chern-Simons action so that the action 
is differentiable with respect to $A_{\tau}^{a}$. 

We make a canonical quantization of the Chern-Simons action $S_{CS}[A]$. 
$S_{CS}[A]$ is invariant under the gauge transformation $\delta
A_{\mu}^{a}= D_{\mu} \lambda^{a}~(D_{\mu}\lambda^{a} = 
\partial_{\mu}\lambda^{a}+\epsilon^{a}{}_{bc}A_{\mu}^{b}\lambda^{c})$ 
and thus describes a
constrained system. The standard recipe of Hamiltonian
formalism of constrained systems yields the following constraints 
in addition to $\pi_{a}^{\tau} = 0$ and the Gauss law constraint:
\footnote{We take $\tau$ as time here. Although, corresponding to 
the radial quantization of conformal field theory on ${\bf R}^{2}$, 
we can formally take $\rho$ as ``time'', the following discussion and 
the final result do not essentially change.}
\begin{eqnarray}
\phi_{a}^{i} \equiv \pi_{a}^{i} + \frac{ik}{8\pi}\epsilon^{ij}
\eta_{ab}A_{j}^{b} &=& 0 \qquad (i,j = \rho, \sigma~~{\rm and}~~
\epsilon^{\rho\sigma}=-1). \label{eq:phi}
\end{eqnarray}
We fix the gauge $A_{\tau} = -i J^{0}$ and will 
explicitly solve the Gauss law constraint later on. 
Thus the Dirac bracket of the remaining dynamical 
variable $A_{i}^{a}$ under the constraint
(\ref{eq:phi}) becomes
\begin{eqnarray}
\left\{A_{i}^{a}(x), A_{j}^{b}(y)\right\}_{D} = \frac{4\pi}{ik}
\epsilon_{ij}\eta^{ab}\delta^{2}(x-y)\qquad
(\epsilon_{\rho\sigma}=+1). \label{eq:Dirac of A}
\end{eqnarray}
By means of the Dirac bracket (\ref{eq:Dirac of A}), the Dirac bracket
between the functionals $F[A]$ and $G[A]$ of the dynamical 
variables $A_{i}^{a}(x)$ is given by
\begin{eqnarray}
\left\{F[A], G[A]\right\}_{D} = \frac{4\pi}{ik}
\epsilon_{ij}\eta^{ab}\int_{\Sigma}\!\!d\rho\wedge d\sigma 
\frac{\delta F[A]}{\delta A_{i}^{a}(x)}
\frac{\delta G[A]}{\delta A_{j}^{b}(x)},
\label{eq:Dirac of function}
\end{eqnarray}
where $\Sigma$ denotes the space-like surface ${\bf R}^{2}$ of
constant $\tau$. 

By using the bracket (\ref{eq:Dirac of function}), the algebra of
the generators of the gauge transformation 
$\delta A_{i}^{a}= D_{i}\lambda^{a}$ can be obtained. The generator of
the gauge transformation is given by the Gauss law constraint:
\begin{eqnarray}
Q(\lambda) = -\frac{ik}{4\pi}\int_{\Sigma}\!\!d\rho\wedge d\sigma~
\eta_{ab} \lambda^{a}F_{\rho\sigma}^{b} + \frac{ik}{4\pi}
\int_{\partial\Sigma}\!\!d\sigma~\eta_{ab}\lambda^{a}A_{\sigma}^{b}, 
\label{eq:generator}
\end{eqnarray}
where $F_{\rho\sigma}^{a}=\partial_{\rho}A_{\sigma}^{a}
-\partial_{\sigma}A_{\rho}^{a}+\epsilon^{a}{}_{bc}A_{\rho}^{b}A_{\sigma}^{c}$. 
$\partial \Sigma$ denotes the boundary of $\Sigma$ at 
$\rho\rightarrow\infty$ and the boundary term is supplemented for
$Q(\lambda)$ to be differentiable with respect to $A_{i}^{a}$.
From (\ref{eq:Dirac of function}) and (\ref{eq:generator}),
one can easily find 
the algebra of the generator $Q(\lambda)$:
\begin{eqnarray}
\left\{Q(\lambda), Q(\eta)\right\}_{D}=
Q\left(\left[\lambda, \eta\right]\right) -
\frac{ik}{4\pi}\int_{\partial \Sigma}\!\!d\sigma~\eta_{ab}
\lambda^{a}\partial_{\sigma}\eta^{b} \qquad \left(\left[\lambda,
\eta\right]^{a} = \epsilon^{a}{}_{bc}\lambda^{b}\eta^{c}\right).
\end{eqnarray}

As mentioned above, we solve explicitly the Gauss law constraint
and the equations of motion of $A_{i}^{a}$ under 
the boundary condition (\ref{eq:boundary condition}) and the gauge
fixing condition for $A_{\tau}^{a}$. The general solutions are given by
\begin{eqnarray}
A_{\tau}=-i J^{0}=\textrm{g}^{-1}\partial_{\tau}~\textrm{g}, \quad 
A_{\bar{z}}=0,\quad {\rm and}\quad 
A_{z}=\textrm{g}^{-1}\hat{A}(z)\textrm{g} 
\qquad(\textrm{g}=e^{-i\tau J^{0}}).
\label{eq:general solution}
\end{eqnarray}
The space of these general solutions is invariant under 
the gauge function $\lambda^{a}=
\textrm{g}^{-1}\hat{\lambda}^{a}(z)\textrm{g}$.
The generator of this gauge transformation gives the global
charge of the Chern-Simons theory on a manifold with the boundary 
\cite{Banados2}.
On this classical phase space, the generator of the gauge
transformation with this gauge function is reduced to
\begin{eqnarray}
Q(\lambda) = \textrm{g}^{-1}\hat{Q}(\hat{\lambda})\textrm{g}~, \qquad
\hat{Q}(\hat{\lambda})=\frac{ik}{4\pi}\oint_{\partial\Sigma}\!\!dz~
\eta_{ab}\hat{\lambda}^{a}(z)\hat{A}^{b}(z).
\end{eqnarray}
We note that the generator is given by the boundary integral only. 
$\hat{Q}(\hat{\lambda})$ satisfies the same algebra as that of
$Q(\lambda)$:
\begin{eqnarray}
\left\{\hat{Q}(\hat{\lambda}), \hat{Q}(\hat{\eta})\right\}_{D}=
\hat{Q}([\hat{\lambda},\hat{\eta}])
-\frac{ik}{4\pi}\oint_{\partial\Sigma}\!\!dz~
\eta_{ab}\hat{\lambda}^{a}\partial_{z}\hat{\eta}^{b}.
\label{eq:alg. of reduced generator}
\end{eqnarray}
In the similar manner to the case of conformal field theory on 
${\bf R}^{2}$, we expand $\hat{A}^{a}(z)$ and $\hat{\lambda}^{a}(z)$
respectively:
\begin{eqnarray}
\hat{A}^{a}(z) = -\frac{2}{k}T^{a}(z) =
-\frac{2}{k}\sum_{n=-\infty}^{\infty}T_{n}^{a}z^{-n-1}, \qquad
\hat{\lambda}^{a}(z)=\sum_{n=-\infty}^{\infty}\hat{\lambda}_{n}^{a}z^{-n}.
\end{eqnarray}
From (\ref{eq:alg. of reduced generator}), the algebra of
the Laurent coefficients of $\hat{A}^{a}(z)$ 
can be obtained as
\begin{eqnarray}
\left\{T_{m}^{a}, T_{n}^{b}\right\}_{D}=\epsilon^{ab}{}_{c}T_{m+n}^{c}
-\frac{k}{2} m~\eta^{ab}\delta_{m+n, 0}.
\label{eq:current algebra}
\end{eqnarray}
This is nothing but the holomorphic part of ${\rm SL}(2;\mathbb{C})$
current algebra. Thus the global charge of ${\rm SL}(2;\mathbb{C})$
Chern-Simons theory on a manifold with the boundary satisfies 
${\rm SL}(2;\mathbb{C})$ current algebra (\ref{eq:current algebra}).   

The solutions (\ref{eq:general solution}), however,
do not satisfy the boundary condition to be asymptotically de Sitter
(\ref{cond2}) in general. Thus we need to require further boundary conditions 
corresponding to (\ref{cond2}) for the Laurent coefficients $T_{m}^{a}$ 
such as\footnote{$T_{m}^{\pm}=T_{m}^{1}\pm i T_{m}^{2}$.}
\begin{eqnarray}
T_{m}^{0}=0, \qquad T_{m}^{+}=-ik \delta_{m+1, 0}.
\end{eqnarray}
One can treat the above conditions as new constraints on $T^{a}(z)$
and obtain the resulting algebra of the remaining variable
$L_{m-1}\equiv -T_{m}^{-}$ by means of the Dirac bracket 
under the new constraints:
\begin{eqnarray}
\left\{L_{m}, L_{n}\right\}_{D}=i (m-n) L_{m+n} -\frac{c}{12}(m^{3}-m)
\delta_{m+n,0}, 
\label{eq:Dirac of Virasoro}
\end{eqnarray}
where $c=6k=\frac{3l}{2G}$.
This is the Virasoro algebra of the asymptotic symmetry on
the asymptotically three-dimensional de Sitter space-time.

However, in order to obtain the corresponding 
quantum algebra acting on the physical Hilbert space, the replacement of 
the Dirac bracket with $-i$ times commutator is required.
Thus the quantum algebra becomes
\begin{eqnarray}
\left[L_{m}, L_{n}\right] = (m-n) L_{m+n} +\frac{i c}{12}
(m^{3}-m)\delta_{m+n,0}.
\end{eqnarray}
As promised, one can find the quantum algebra of the asymptotic
symmetry is the Virasoro algebra with a {\it pure imaginary} central 
charge $i c$. 

Finally we comment on the relation between this algebra and the Virasoro
algebra discussed in Section 3. From the relation $T^{-}(z) = -i b(z)$
(see (\ref{general-solution})), $T^{-}_{m} = -i\tilde{L}_{m-1}$ leads to
the corresponding quantum algebra:
\begin{eqnarray}
\left[\tilde{L}_{m}, \tilde{L}_{n}\right] = -i (m-n) \tilde{L}_{m+n} - 
\frac{i c}{12}(m^{3}-m)\delta_{m+n,0}.
\label{eq:algebra of b}    
\end{eqnarray}
The central extension term of the algebra in terms of $\tilde{L}_{m}$
still remains {\it pure imaginary}.
Two different algebras, (\ref{quantum-Virasoro}) 
and (\ref{eq:algebra of b}), are based on
the Poisson bracket of the same reduced variables. 
This discrepancy originates from the different choice of
the symplectic structure which defines the Poisson bracket 
between the reduced variables.

%%% References %%%

\end{document}